# Two Trends in Mobile Security: Financial Motives and Transitioning from Static to Dynamic Analysis

Emre Erturk
School of Computing
Eastern Institute of Technology
Napier, New Zealand
eerturk@eit.ac.nz

*Abstract*—**The goal of this paper is to analyze the behavior and intent of recent types of privacy-invasive Android adware. There are two recent trends in this area: more financial (rather than ego) motives, and the development of more dynamic analysis tools. This paper starts with a review of Android mobile operating system security, and also addresses the pros and cons of open source operating system security. Static analysis of malware provides high quality results and leads to a good understanding as shown in this paper. However, as malware grows in number and complexity, there have been recent efforts to automate the detection mechanisms and many of the static tasks. As Android's market share is rapidly growing around the world, Android security will be a crucial area of research for IT security professionals and their academic counterparts. The upside of the current situation is that malware is being quickly exposed, thanks to open-source software development tools. This cooperation is important in curbing the widespread theft of personal information with monetary value.**

## I. INTRODUCTION

The goal of this paper is to analyze the behavior and intent of recent types of privacy-invasive Android adware. This paper starts with a review of recent developments in Android mobile operating system security. Many tools and applications used either to create malware or to clean malware are open-source and free. The scope of the analytical part of this paper includes different samples of the so-called Plankton malware, other similar but recent adware that have followed in Plankton's footsteps, a sample of drive-by malware, and the Tigerbot Trojan (to set a contrast). The aspiration of adware perpetrators seems to be to spread in the highest numbers possible, just like ocean plankton existing in extremely large numbers. The financial motives behind their activities are clear. Their objective is not to disrupt computer users and companies but to make quick financial gains. In looking at adware, this paper also addresses ads in general and ad networks. As a result of a static analysis of sample mobile threats, this paper also addresses the pros and cons of open source operating system security, and dynamic analysis tools and projects. This study concludes by tying the review and analytical findings to practical recommendations.

## II. REVIEW OF MOBILE SECURITY LITERATURE

So far mobile malware has largely targeted the Android operating system as opposed to other systems, for example, iOS and Blackberry OS. Some of the recent threats have been elevated in the media. Skeptics may attribute this partly to computer security software companies wanting to create a new market on mobile systems for their products; similar alerts were published several years ago regarding the other major mobile operating systems even though the long run outcomes have not turned out to be grave. One difference this time is that the target is a semi-open source operating system. Researchers continue to discover new vulnerabilities on Android devices. Serious examples include botnet malware involving numerous apps downloaded by a great number of users from the Android Market. In recent years, many file sharing, peer-to-peer, and torrent web sites have been targeted or stopped by government authorities and copyright owners' associations around the world. This trend may lead to fewer Android malware and adware as some of these web sites distributed Android software (containing malware and adware) and operated ad networks.

A service called Bouncer scans the Android Market (Google Play) for known malware and simulates newly added apps to catch the misbehaving ones. Bouncer has its shortcomings as a dynamic tool because it can be circumvented if infected applications are intelligent enough to behave normally during the scan and then misbehave in a normal environment. Bouncer checks an app only for five minutes, using the same the phone account with exactly one phone contact and two photos [2]. Although Google has a track record of removing reported misbehaving apps, during the course of writing this paper, there have been cases observed where some apps either remained on or came back to Google Play.

According to Google, Android included certain core security features from the start, including sandboxing (putting virtual walls between applications and other software on the device) and using a permission system, which shows the user during installation what types of access each new app demands. Furthermore, Android's recent version 4.1 provides greater user control and security against unwanted notifications displayed on the smartphone. Users can click and see exactly which apps have generated any of the notifications, and choose to turn them off or uninstall the application.

In addition, behavioral biometric security will play a larger role with Android's recent versions 4 and 4.1 as they come with a facial gesture recognition app for security (which is enhanced with a blink option so that it cannot be tricked by a photograph). Other examples of behavioral security are new software products, which understand and verify each unique user's keystroke and user interface patterns. These may protect important data, if a device were stolen or compromised.

The question of whether open source software is more secure than proprietary software has been discussed for a long time. Although some of this discussion is technical and empirical, it is also inevitably influenced by the authors' own background and even stereotypes. Similar to Linux, the Android brand has a socially benevolent image as a virtue of having open-source code (though the hardware manufacture and much of the software development is for profit). This lends sympathy in society in general and among computer enthusiasts toward the cute little robot (the Android logo). As it is costless and easy to access, many organizations and consumers have become prone to blindly trusting open-source software [14]. On the other hand, open source communities have the potential to address problems faster than commercial application teams.

### III. STATIC VS DYNAMIC ANALYSIS

There are two ways to analyze malware in general whether desktop or mobile: dynamic analysis and static analysis. Dynamic analysis involves automated tools to execute the malware in a controlled system environment and check for malicious patterns. Therefore a large sample of malware can studied quickly. On the other hand, dynamic analysis is not yet common for mobile devices because of the difficulty of replicating the numerous mobile hardware, mobile operating systems, and their many different versions.

An advantage of Android is the ability to install apps through different methods, for example, direct installation from an app market, downloading and manually installing (Android Packages (APK file format), or Java MIDlets (a certain type of platform independent mobile apps). However, this sometimes means greater potential for malware infection. Mobile malware have different ways of spreading as opposed to desktop malware, which further makes dynamic analysis more difficult [11]. These spreading vectors include SMS/MMS (Short Message Service/ Multimedia Messaging Service) messages with links, infected applications in official and unofficial markets, seemingly normal applications with a malware installer hidden within the installation APK file, wireless connections and drive-by infection through a compromised web site. Obfuscation and encryption are quite common for mobile malware, which also require extra steps in the analysis [10]. Other sophisticated techniques used for evasion include polymorphism (changing variables and files), and steganography (hiding information in unexpected places) [13].

Static analysis, which involves human work to trace and study malware, is more common because of the relatively small number of mobile malware. Static analysis can provide higher quality results and lead to a better understanding. This approach has been used in the analytical part of this paper.

Android applications come as compressed packages (APK). Every application package contains standard components. Using certain tools, compiled byte code is reverse-engineered into human readable format. The first step is to obtain malware sample from a repository shared between researchers and professionals. The individual pieces of each package can be extracted with a zip utility such as 7-Zip. The first component to look at is the AndroidManifest.xml file. This initially comes in binary xml format, and contains the system permissions to be granted to the app. With the EditIX utility, it is possible to convert and read this information in clear text. The second and biggest component is classes.dex file, which contains the program code, and thus the main or initial payload. This file is first converted to a Java archive with the Dex2Jar utility. Then it is legible with a Java editor such as JD-GUI. The other interesting component in each package is the resources.arsc file, which may provide further clues to the analyst.

The payload of Plankton adware consists of executable Java byte code that is initiated in the background. After collecting information about the infected device, Plankton contacts a specific web site and is capable of downloading further payload. It also allows commands to be given remotely from the contacted site. On a positive note, Plankton is currently detectable by mobile security software and by ad network detector software. While writing this paper, the actual Java code of the sample malware was analyzed. The list of commands that could be given was made legible through the Java de-compiler. Furthermore, a number of tactics were observed. Deprecated and obsolete code was left in the program order to obscure the active code. Similarly, programming interfaces were named in meaningless ways to either divert or confuse anyone who might be reading the code. The web address of the contacted remote site is visible in one of the files within the package but again the ploy was to put this information in a separate and less expected place.

First samples of Plankton were uncovered in 2011 by North Carolina State University researchers [11]. Samples apps containing Plankton were available in 2012 on official and secondary markets while writing this paper and one such game app was installed on a test smartphone. Ads were frequently displayed on the infected device (in and outside of the game) along with suggestions for other games to download. The browser home page was changed, and a shortcut was placed on the desktop. The second sample of Plankton involved a game with pictures of pretty women, a common social engineering exploitation technique. Currently Plankton can be seen as part of an aggressive ad network rather than a Trojan.

The group of Android malware that includes Spy-E and SNDAPPS displays unsolicited advertisements, generates notifications, gathers user and phone information, and communicates with an ad web site. If the variants NickiSpy and SMS.Boxer are considered as related, their payload includes draining money by sending SMS from the infected device to a premium number [15]. In contrast, the new drive-by malware is downloaded by surfing a malicious web site instead of using apps. It is also different since it does not engage in any of the above activities. It may be a reconnaissance tool to reach mobile devices (which may be attached to corporate networks) and possibly try to use them as a jumping board for an attack.

One of the newcomers among Android malware in April 2012 was TigerBot. This malware allows remote access by the attacker and can be controlled via SMS messages [18]. It will listen for specific messages, which can steal contacts lists and screenshots, change network settings, deactivate other software, and control running processes. During the static analysis in this paper, the apps containing this malware have been shown to obtain (if installed) an unusual variety of intrusive permissions. TigerBot is beyond adware, actually a Trojan going so far as to disguise itself with a Google icon.

The Honeynet online community has been at the forefront in showcasing new dynamic analysis tools for Android malware [19]. These tools include DroidBox and APKInspector, which have been trialed during the course of writing this paper. In some ways, these tools do not completely automate all human analysis. Rather, they synergistically integrate individual tools into a suite, run them together with less effort, and produce visual graphs, and work flows that greatly assist the examination of malware. Technically, these tools can be used broadly to help analyze any Android software. However, there are features that make them especially geared toward catching malicious activity, for example screen tabs for permissions and phone calls, and measuring SMS, cryptographic activity, and data leaks.

A honeypot is a computer system that can be used as a trap and be monitored in order to detect and study new attacks. This type of system can help automate security analysis by reducing the manual search for malicious apps and servers, and by capturing live information and audit logging the attacks. HoneyDroid started in 2011 as the first effort to build a honeypot system for the Android platform [20]. Although honeypot ideas were explored in earlier years with Windows Mobile and Symbian, the projects have not matured or stayed on as a result of those platforms being phased out or revamped. The first challenge is making the Android honeypot visible to attackers. Perhaps a main reason that undermines the feasibility of a honeypot system is that smartphones rarely run network services to be noticed by itself on the internet and an active seeking of malicious sites or people would somewhat conflict with the idea of a honeypot itself. The second challenge is that the core components of the honeypot system must themselves not be vulnerable as that would cause them to not function properly for detecting and recording malicious behavior. HoneyDroid's solution to this dilemma is to virtually run Android on top of another securely isolated microkernel operating system, all residing on real phone hardware [20]. The drawback of this approach is that virtualization may be noticed by malware or it may not run the same way in the honeypot.

IV. FINANCIAL MOTIVES

It is not just the core technical attributes of an operating system that determines its exposure to risk; the greater popularity of any operating system platform will lead to more attempts by perpetrators to target that platform and its user base. Nevertheless, obscurity does not mean better security; in comparison to proprietary operating systems, open-source operating systems allow and encourage a greater number of people to work against malware. The body of knowledge suggests that hackers typically do not go about finding vulnerabilities by reading the underlying source code; they do so by probing and trying different tactics from the outside [5]. As a result, brute force attacks are often used. Mike Calce is a famous former hacker, and currently a consultant and the author of a book on internet security. In an interview in 2012, he also stated his belief that the ulterior motive for most of today's hackers is monetary gain [7]. In addition, the risk is shifting more and more from governments and companies to individuals.

Adware is any software package that automatically presents advertisements to users by guessing from their previous surfing or search activities. This involves collecting information, often by user consent, but in some cases, stealing important personal information for ulterior harmful motives. Aside from adware used intentionally by an ad network, other intrusive adware may also exploit an ad network and subvert revenue and information from the owners of the ad network [3]. Social engineering is the art of manipulating people using trickery or deception for the purpose of information gathering, fraud, or system access. Phishing is a common type of social engineering where the attacker notifies users that they need to take action. The email or pop-up contains a link to a fake web site for collecting the user's id and password. Adware may also be used in combination with phishing or automated click fraud.

Some of the pushed ads are displayed through the Android notification bar. The developers of an app can earn part of the advertising revenues this way. This is usually done via cloud messaging, where the server can send notifications to a smartphone without the device requesting them first [6]. This of course requires that the app has gained the necessary permissions to auto-start at boot and run in the background. This may cause drain of resources such as the battery, temporary files and images occupying the device's storage space, and increased internet data usage and roaming charges [6]. Furthermore, if the notifications are clicked on, they may lead to phishing, hacking, or other high risk web sites. Google has updated Google Play Developer Program Policies in August 2012 to prohibit inappropriate advertising activities. With these policies, aside from generally illegal and offensive content, Google is also banning deceptive adware behavior in apps such as impersonating the operating system, making changes to the user's device, hiding from the user which app is generating the ads, and not giving the user the ability to adjust advertising preferences [1].

The analysis of malware and recommendations against them are not based only on program logic because there are supplementary geographic and financial trends. Malware and adware can be better understood in the context of their monetary objectives and countries where they often originate and are distributed (in particular China and Russia). For example, one of the most recent and sophisticated malware that downloads paid apps and media files, leading to unwanted monetary charges, is based in China. The malware called TROJMMARKETPLAY (discovered by security vendors) comes in multiple versions, some of which even includes experimental code. It changes the smartphone's access point name, connects automatically to a secondary app store, closes normal consent windows, and intercepts verification codes messages so that the user remains unaware in the interim [9].

Secondary (unofficial) app markets, where many malware have been found, seem to have grown also as a result of language factors in the case of China, Hong Kong, and other parts of East Asia. The official Android Market (Google Play) is still blocked as a result of government restrictions in China. This has given rise to many secondary local app markets there.

On one hand, many free apps rely on advertising to support their development. On the other hand, as can be seen from the sample malware, certain apps have crossed the line from merely displaying ads to pushing (or forcing) products to the user, harvesting private data for future use (e.g. spam or other use), and even extracting fraudulent revenues. It is possible for hackers to rent premium rate numbers anonymously (for generating dialing or SMS fraud) in Russia and other Eastern European countries whereas this is not possible in many other countries [17]. This type of fraud affecting Russian Android users goes back to 2010, involved sending SMS to certain numbers that cost the users US$ 5 per message [11]. A recent example of the same SMS fraud activity involved fake Skype apps that were downloaded through Russian web sites as Java MIDlets, which again cause monetary damage [8].

Recent types of Android malware resemble their desktop-based predecessors rather than being genuinely created for a specific operating system. Therefore it is necessary to recall the key motives of the hacker subculture in general that also pertain to malicious Android activity. These are entertainment, ego, status, entrance to a social group, money, and cause [4]. Money, a less common motivator in the 1980s, has grown as a result of the World Wide Web, the enormous volume of commercial transactions, and the vast amount of personal information available and exchanged online [4]. The stolen information (credit cards, bank accounts, logins, etc.) is sold between hackers worldwide. Malware and botnets (collection of compromised computers) are also traded in this underground economy. This black market allows skilled hackers to make a profit by selling their expertise and spoils to others [4].

## V. CONCLUSION

Android is presumably the most popular mobile operating system in most countries (including high-income countries). Android has achieved the market breakthrough that the proponents of open-source and Linux software have been waiting for, in economically developing countries as well [12]. Android has been successfully adopted by many hardware manufacturers, with a wide range of expensive and low priced models. By 2015, low-end Android smartphones are expected by market researchers to seize 80% of the market in Africa, India, and China [16]. This great market share across the world also makes Android vulnerable as it provides a large financial incentive for hackers and malware perpetrators to target its individual users and their private information.

Android security will be a crucial area of research for IT security professionals and their academic counterparts. The upside of the current situation is that malware is being quickly disclosed, thanks to accessible and open-source software development tools. Open source software facilitates worldwide community response to security threats. Cooperation against malware needs to increase, not just within individual countries, but across different geographic regions of the world.

In the future, honeypots and dynamic analysis tool suites should put more emphasis on detecting and understanding malware behavior that may have monetary consequences. In order for honeypots to become more visible and successful, they should be deployed in different parts of the world and be capable of operating in different languages. The future holds promise for interesting developments in smartphone security.